\documentclass[sigconf,authorversion,nonacm]{acmart}

\usepackage{verbatim}
\usepackage{multirow}
\usepackage{listings}

\AtBeginDocument{%
  \providecommand\BibTeX{{%
    \normalfont B\kern-0.5em{\scshape i\kern-0.25em b}\kern-0.8em\TeX}}}

\begin{document}

\title{Transformer-based Vulnerability Detection in Code at EditTime: Zero-shot, Few-shot, or Fine-tuning?}

\author[1]{Aaron Chan, Anant Kharkar, Roshanak Zilouchian Moghaddam, Yevhen Mohylevskyy, \\Alec Helyar, Eslam Kamal, Mohamed Elkamhawy, Neel Sundaresan} 

\affiliation{ \institution{Microsoft}
 \city{Redmond}
 \country{USA}}


\begin{abstract}
Software vulnerabilities bear enterprises significant costs. Despite extensive efforts in research and development of software vulnerability detection methods, uncaught vulnerabilities continue to put software owners and users at risk. Many current vulnerability detection methods require that code snippets can compile and build before attempting detection. This, unfortunately, introduces a long latency between the time a vulnerability is injected to the time it is removed, which can substantially increases the cost of fixing a vulnerability. We recognize that the current advances in machine learning can be used to detect vulnerable code patterns on syntactically incomplete code snippets as the developer is writing the code at \textit{EditTime}. In this paper we present a practical system that leverages deep learning on a large-scale data set of vulnerable code patterns to learn complex manifestations of more than 250 vulnerability types and detect vulnerable code patterns at \textit{EditTime}. We discuss zero-shot, few-shot, and fine-tuning approaches on state of the art pre-trained Large Language Models (LLMs). We show that in comparison with state of the art vulnerability detection models our approach improves the state of the art by 10\%. We also evaluate our approach to detect vulnerability in auto-generated code by code LLMs. Evaluation on a benchmark of high-risk code scenarios shows a reduction of up to 90\% vulnerability reduction.

\end{abstract}

\keywords{Transformers, Software Vulnerabilities, Vulnerability Detection}


\maketitle
\pagestyle{plain}
\section{Introduction}
Despite development of many tools and best practices \cite{mssdl, ardi2007can, cmucert, taylor2008moving}, uncaught vulnerabilities persist in code\cite{vuldb, cwe} and impact users and cost companies time and money\cite{vulexample}. In recent years, many approaches have been developed to detect vulnerabilities in software. These approaches range from traditional dynamic analysis \cite{li2019cerebro, li2017steelix, wang2019superion}, rule-based static analysis \cite{xu2017spain, chandramohan2016bingo}, feature-based machine learning solutions \cite{neuhaus2007predicting}, to more recent deep learning based solutions \cite{zhou2019devign, li2018vuldeepecker, russell2018automated, chakraborty2021deep}.

Approaches that rely on dynamic analysis often suffer from low code coverage issues. The rule-based static analysis approaches typically involve manual efforts by experts to characterize and add new or evolved vulnerability patterns on a regular basis. Similarly, feature based machine learning approaches rely on hand-crafted features by human experts. 
More recently, deep learning-based approaches have emerged as an alternative which can learn vulnerability patterns automatically without expert involvement. However the majority of these Deep learning approaches require a complete source file \cite{peng2015building}, a complete function \cite{le2019maximal, lee2017learning, russell2018automated}, or a complete code statement \cite{sestili2018towards, choi2017end} to analyze.  
As a result, to the best of our knowledge, the existing deep learning approaches cannot find vulnerabilities at \textit{EditTime} while the developer is typing and the code is not yet syntactically correct or complete. 

\begin{figure}[ht]
    \includegraphics[scale=0.5]{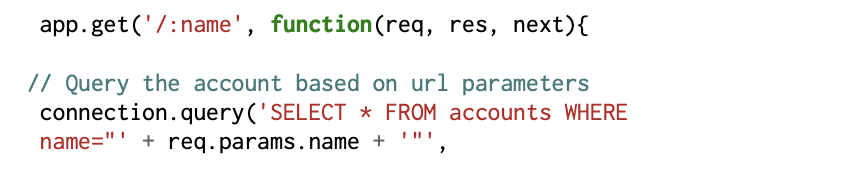}
    \caption{The best time to notify the developer about a \emph{SQL-Injection} vulnerability is at \textit{EditTime}, right after the developer has made the mistake.}
    \label{figure:sample-vuln}
    \end{figure}



The literature indicates that the cost of fixing a fault is positively correlated with the fault ignorance time (i.e the time when a vulnerability is injected to the time it is removed)\cite{baziuk1995bnr, barry1981software, humphrey1995discipline}. Therefore, the developer should be notified about the fault in a timely manner, when the fault is relevant to the current programming task\cite{layman2007toward}. For example, in the code snippet shown in Figure~\ref{figure:sample-vuln}, the best time to notify the developer about the SQL injection they have introduced is right at the end of the query, even though their code is not syntactically complete. 
The importance of interactively identifying bugs and vulnerabilities in code at \textit{EditTime} has been recognized by prior work \cite{xie2011aside}. Our work extends this line of prior work by leveraging new advancements in deep learning to detect a variety of vulnerabilities in incomplete code snippets as the developer is typing. 

To build our vulnerability detection solution, we collect more than 500K vulnerable code snippets from running static analyzers on open source repositories. We then develop six model variations using common learning approaches for pre-trained LLMs: zero-shot, few-shot and fine-tuning. We apply these learning techniques on CodeBERT\cite{codebert} and two state of the art pre-trained LLMs provided by OpenAI: \texttt{code-davinci-002}, \texttt{text-davinci-003}. Our comparative evaluation of these variations shows that fine-tuning CodeBERT offers the best balance between precision and recall (precision of 59\% and recall of 63\%). However, both zero-shot and few-shot learning on recent \texttt{text-davinci-003}, which is an InstructGPT\cite{ouyang2022training} based language model, offer better recall (78\% and 75\%) and slightly lower precision (47\% and 49\%). We discuss the benefits and challenges of each learning approach in terms of model size and performance.

We conduct two experiments using our best performing model. The first experiment compares this model with existing vulnerability detection approaches on four established benchmark datasets. We show that, in comparison with existing approaches, our approach improves recall up to 10\% and precision up to 8\%. 

The second experiment studies the generalizability of our approach beyond manual code edits by developers, where we investigate the effectiveness of our approach in detecting vulnerabilities in automated code edits. With the recent wide-spread adoption of code completion tools \cite{Copilot, svyatkovskiy2020intellicode, tabnine}, many future code edits are expected to be automatically generated by code LLMs. 
Prior work suggests that code LLMs make similar mistakes to developers \cite{codex_apr}, thus auto-generated code edits can have potentially severe security concerns. For example, in certain scenarios, up to 40\% of the completions generated by code LLMs included vulnerable code patterns \cite{nyu_benchmark}. 
We evaluated our vulnerability detection model on a variant of the benchmark introduced by Pearce et al. \cite{nyu_benchmark}. Our evaluation shows that using our model yields 90\% reduction in vulnerability rate.  

Finally we discuss the deployment journey of our vulnerability detection model in a VSCode extension which has resulted in 80\% reduction in rate of vulnerability in code edited with the extension. 

This paper has the following contributions:
\begin{enumerate}
    \item we present a production-quality vulnerability detection system which can identify vulnerabilities in incomplete code snippets on the order of milliseconds and therefore serves developers at \textbf{EditTime},
    \item we explore and discuss the benefits and challenges of three common learning approaches for pre-trained LLMs on the task of vulnerability detection: zero-shot, few-shot, and fine-tuning,
    \item we expand the benchmark introduced by Pearce et al. and make it available to the code vulnerability detection community as a future \textit{EditTime} benchmark.
\end{enumerate}

\section{Related Work}
We discuss how our work builds upon and extends prior work in the area of Vulnerability Detection, Deep Learning methods for Vulnerability Detection, and Vulnerability Detection in auto-generated code. 

\subsection{Vulnerability Detection}
There are two modalities of analyzing source code to search for vulnerabilities: dynamic analysis, in which a code snippet is executed and the stack trace examined for signatures of a specific vulnerability; and static analysis, in which the source code is analyzed without execution. One common method of dynamic analysis is fuzzing, in which a program is run on many different inputs to explore the space of program execution traces. Optimization techniques to efficiently search this program space are an active area of research \cite{confetti, fuzz_pathtran, fuzz_temporal}. However, fuzzing large code-bases by executing arbitrary code is infeasible, and has led to the development of static analysis techniques. Static analyzers search code-bases for semantic patterns of bugs or vulnerabilities\cite{owasp}. CodeQL \cite{codeqlref} is a popular scalable static analyzer for such pattern-based searches. CodeQL offers a query language to write rule-based queries for specific anti-patterns. We build upon this line of research by developing a vulnerability detection model that leverages state of the art deep learning approaches and therefore is capable of learning a variety of vulnerability types without involving human experts. 

\subsection{Deep Learning Vulnerability Detection}
Since natural language and programming languages both possess sequential structure, model architectures with success in NLP, such as GRU, LSTM, and Transformer, have also shown promise in vulnerability detection \cite{selfrep_autovuln}. One unique characteristic of source code, not shared by natural language, is the inherent graph structure of programs. VulDeePecker \cite{li2018vuldeepecker} extracts code gadgets from data dependency graphs and trains a BiLSTM classifier to detect vulnerabilities in a dataset of two different vulnerability types. SySeVR \cite{li2021sysevr} builds on VulDeePecker by using both syntactic and semantic information to train vulnerability classifiers on a dataset of 126 vulnerability types. Other works leverage graph neural networks to supplement language models: IVdetect \cite{ivdetect} uses GRU embeddings with a Feature-Attention Graph Convolutional Network (FA-GCN) to detect vulnerabilities in program dependence graphs. LineVul \cite{linevul} extends IVdetect by leveraging pretrained CodeBERT \cite{codebert} to perform line-level vulnerability localization in addition to method-level detection. Devign \cite{zhou2019devign} learns embeddings of code representation graphs and trains a gated graph recurrent network to detect vulnerabilities for graph-level predictions. It evaluates this model on vulnerabilities from FFmpeg and QEMU.
ReVeal \cite{chakraborty2021deep} introduces a dataset of vulnerabilities from real-world software projects as a benchmark for vulnerability detection models. It demonstrates this benchmark using a gated graph neural network (GGNN) trained on code property graphs. Our research extends prior work on vulnerability detection by developing a neural vulnerability detection model for seven languages: JavaScript, Python, Go, Java, C++, C\# and Ruby that is capable of detecting vulnerabilities at \textit{EditTime}.

\subsection{Vulnerability Detection in Auto-generated Code}
Transformer-based models trained on source code have achieved state of the art results for code generation\cite{codexref, nijkamp2022conversational, li2022competition}. These models have unprecedented utility as developer-assistance tools, and are being deployed into products \cite{Copilot, svyatkovskiy2020intellicode, tabnine}. Since these code LLMs have been trained on large datasets of human-written code, their code outputs are increasingly likely to follow the patterns (or anti-patterns) of human developers. For example, \cite{codex_apr} evaluated Codex on Java LeetCode questions and found that functionally incorrect completions from Codex share similar anti-patterns to incorrect code written by humans. Similarly, \cite{nyu_benchmark} examined Codex with regard to generating vulnerabilities and found that, in certain scenarios, Codex did indeed generate vulnerable code patterns. In another study, \cite{siddiq2022empirical} showed that during functionality tests with the HumanEval dataset, GitHub Copilot generates certain CWEs in around 2\% of cases. These studies suggest that code LLMs are likely learning to generate both bad and good code. Luckily, since vulnerable code patterns are rare in human code and therefore make up a small portion of training data, they are also rare in completions from Code LLMs. For example, a recent work studying the effect of code LLM assistance did not find any conclusive evidence that code LLMs are more likely to generate vulnerabilities than humans \cite{sandoval2022security}. Nevertheless, given the high cost associated with code vulnerabilities, reducing vulnerable code patterns generated by code LLMs is essential in ensuring that their prolonged usage is safe, especially for novice programmers. To this end, our research yields a model that can operate on syntactically incomplete code, and therefore detect vulnerabilities in auto-generated code snippets as well as the developer written code snippets at \textit{EditTime}.

\section{Detecting Vulnerabilities at \textit{EditTime}} 
To detect vulnerabilities at \textit{EditTime} we develop six model variations using common learning approaches for pre-trained LLMs: zero-shot, few-shot and fine-tuning. Below, we first explain the training data we collected for our fine-tuning process. We then explain the model variants and their respective performances.

\subsection{Data Collection}
Our dataset consists of vulnerable code patterns detected in public GitHub repositories by the GitHub LGTM service. GitHub LGTM service runs CodeQL \cite{codeqlref}, a scalable static analyzer, on public GitHub repositories to identify a variety of vulnerable code patterns, including hard-coded credentials, SQL injections, and path injections. In this work, we select a subset of detected CodeQL issues that correspond to a set of ``Common Weakness Enumeration" (CWE) \cite{mitre} in each of seven languages: JavaScript, Python, Go, Java, C++, C\#, and Ruby.\footnote{The source code for all CodeQL queries we ran can be found at https://github.com/github/codeql under the ql/Security folders.} Table \ref{tab:data-vul-summary} shows the summary statistics of the curated dataset.

\begin{table}[h]
    \caption{Summary statistics of vulnerability detection training dataset}
    \centering
    \begin{tabular}{c|c|c|c}
        Model & Coverage & Vulnerable & Non Vulnerable \\
        \hline 
        Javascript & 70 CWEs & 266,342 & 2,293,712 \\
        Python & 37 CWEs & 149,158 & 1,493,972 \\
        Go & 29 CWEs & 50,233 & 535,180 \\
        Java & 44 CWEs & 33,485 & 431,726 \\
        C++ & 32 CWEs & 7,222 & 215,722\\
        C\# & 54 CWEs & 3,341 & 27,731 \\
        Ruby & 19 CWEs & 137 & 1,957 \\
    \end{tabular}
    \label{tab:data-vul-summary}
\end{table}

Each detected vulnerability contains the following:
\begin{itemize}
    \item \emph{Vulnerability title:} the type of detected vulnerability, corresponding to a CWE category
    \item \emph{Message:} a detailed error message explaining the detected vulnerability
    \item \emph{Location:} file path, line, and column where the issue starts and ends
\end{itemize}

\subsection{Data Pre-Processing}
\begin{figure}[ht]
    \includegraphics[scale=0.5]{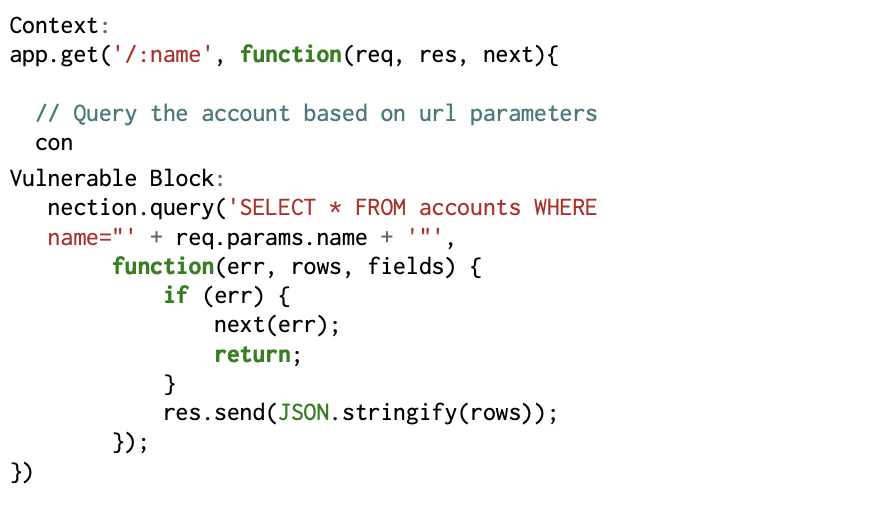}
    \caption{A sample context and vulnerable block from our training data.  In this example, the vulnerable block contains the \emph{SQL-Injection} vulnerability.}
    \label{figure:sample-prompt}
    \end{figure}

Given the data we collected, the goal of our data pre-processing step was to synthesize the following triplets: $(c_i, v_i, l_i)$ where $c$ is a snippet of code that the model takes in as context, $v$ is the block of code that is vulnerable, and $l$ is the label for the vulnerability type. Our process for synthesizing training data is as follows: we first collect files with identified vulnerabilities from all 8,815 repositories in which these issues were detected. For each file, we extract the Abstract Syntax Tree (AST), and search for nodes of the tree that contain complete scopes. Specifically, we look for statements (such as \texttt{if} and \texttt{export} statements), methods, declarations and clauses. For these blocks of code, if the code block contains a vulnerable code pattern, we randomly split the block at some character before the start of the detected vulnerability. Otherwise, if no vulnerability is detected, we randomly split the block at any point. This splitting process mimics a possible code state at \textit{EditTime}. Therefore, models trained on this data should be able to detect a vulnerable block in a syntactically incorrect and incomplete code snippet. 
The first segment of the block is labeled as the ``context" and the second segment as the ``vulnerable block". Figure \ref{figure:sample-prompt} shows an example of a context and vulnerable block pair. If the completion contains a vulnerability, the corresponding vulnerability type is assigned to the example as a label. This separation between context and vulnerable block forces the model to focus on the vulnerable block given the context. We used a 85/5/10 train-validation-test split at the repository level, such that the training set is comprised of 85\% of the repositories, the validation set is comprised of 5\% of the repositories and the test set is comprised of 10\% of the repositories. We then removed any example in the training set that matches a test set example.

\subsection{Models}
In order to detect vulnerabilities at \textit{EditTime}, we develop six model variations using three learning approaches for LLMs: zero-shot learning, few-shot learning, and fine-tuning. We use these learning approaches on three pre-trained LLMs as our base:
\begin{itemize}
    \item \texttt{code-davinci-002}: full-size Codex model which is trained on source code from GitHub.
    \item \texttt{text-davinci-003}: Codex model based on InstructGPT \cite{instructgpt}, using reinforcement learning from human feedback (RLHF).
    \item CodeBERT: transformer pre-trained trained on bimodal data (code and documents)\cite{codebert}. Its architecture follows RoBERTa-base \cite{DevlinCLT19}, with 12 layers, 12 attention heads, and a hidden size of 768. 
\end{itemize}
Using three learning approaches on the above pre-trained models yields six models that are listed in Table \ref{table:github_cwes_results}. We explain the details of developing these models below. 

\subsubsection{Zero-shot Learning}
In zero-shot learning, we provide a pre-trained model with a prompt that specifies our desired output. In our case, we prompt \texttt{code\-davinci-002} and \texttt{text-davinci-003}. In order to obtain a suitable prompt for asking each model to detect vulnerabilities, we leverage the model's own definition of vulnerability detection. We first prompt the model to define the vulnerability detection task and use this task description in the final prompt for the task. 

\begin{figure}[ht]
    \includegraphics[scale=0.5]{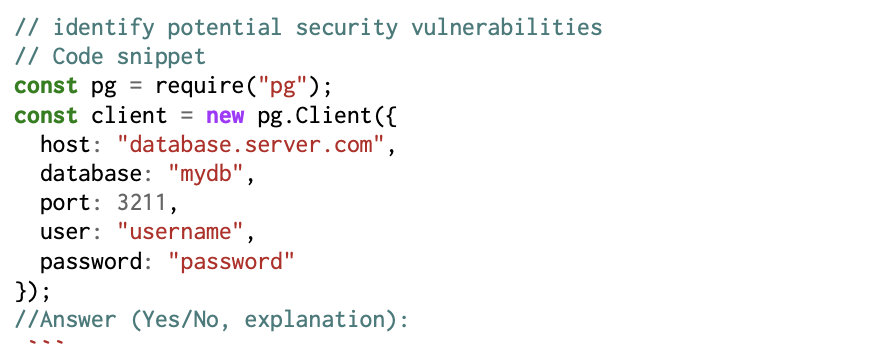}
    \caption{A sample prompt created based on our template for zero-shot setting}
    \label{fig:prompt-template1}
    \end{figure}

For \texttt{code-davinci-002} we prompt the model with \textit{"\# We run CodeQL security queries in order to "} on temperature 0.8 for four times and pick the top results. This yields the following prompt variations: \textit{"identify potential security vulnerabilities"}, \textit{"find potential security issues"}, \textit{"find security vulnerabilities"}, and \textit{"detect security vulnerabilities"}. We place these variations in a zero-shot prompt following the below template:\\
\begin{lstlisting}[linewidth=6cm, basicstyle=\small\ttfamily, numbers=none, xleftmargin=.01\textwidth, commentstyle=\color{dkgreen}
]
    <phrase>
    <comment> Code snippet
    <code snippet>
    <comment> Answer (Yes/No, explanation):
\end{lstlisting}
Here $<phrase>$ corresponds to one of the prompt phrasing variations above, $<comment>$ refers to a language specific comment indicator (e.g. "\#" for Python) and $<code$ $snippet>$ refers to the code snippet in question. Figure \ref{fig:prompt-template1} shows a sample prompt following this template. In response to this prompt, the model outputs either Yes or No, followed by an explanation which can be used for observations and debugging purposes. We check for the presence of "Yes" or "No" to determine the model's decision. We refer to this zero-shot approach as \textbf{CodexZero}.

Similarly, for \texttt{text-davinci-003}, we first prompt the model four times at temperature 0.8 with the question, \textit{"What would you accomplish by running CodeQL security queries?"}. We then ask the model to rephrase its response four times. This yields the following unique phrases: \textit{"identify potential security vulnerabilities"}, \textit{"spot any security weaknesses"}, \textit{"detect any security risks"}, \textit{"determine any security issues"}. We then try these variations in a zero-shot prompt following a similar template: 
\begin{lstlisting}[linewidth=6cm, basicstyle=\small\ttfamily, numbers=none, xleftmargin=.01\textwidth, commentstyle=\color{dkgreen}
]
    <phrase>
    <code snippet>
    Answer (Yes/No):
\end{lstlisting}
We check for the presence of "Yes" or "No" in the model's response to determine the model's decision. We refer to this zero-shot approach as \textbf{TextZero}.

\subsubsection{Few-shot Learning}
Few-shot learning builds on zero-shot learning by providing example input-output pairs, in addition to the zero-shot prompt. For our study, we utilize the best performing prompt variations of \texttt{code-davinci-002} and \texttt{text\-davinci-003} from our zero-shot learning in the same template format. We then prepend additional examples in the same template format before finally inserting the code snippet of interest and prompting the model for the answer.

To create the examples, we prompt each model with the phrase, \textit{"Provide an example in <Language> of a code snippet that contains <Vulnerability Name> security vulnerability. Output the code only, do not include text:"} for each of the languages and types of vulnerabilities in Table \ref{table:github_cwes}. We prompt with this template three times for each vulnerability and language pair, yielding 150 vulnerable examples. We then prompt the model with \textit{"Provide an example in <Language> of a code snippet. Output the code only, do not include text:"}, to retrieve non vulnerable samples of code. We manually evaluate each sample to ensure that they were vulnerable or non vulnerable as intended. In total, there were 297 examples. We refer to the models resulting from the above few-shot learning process with \texttt{code-davinci-002} and \texttt{text-davinci-003} as \textbf{CodexFew} and \textbf{TextFew}, respectively.

\subsubsection{Fine-tuning}
For fine-tuning, we focus on pre-trained code LLMs only and fine-tune \texttt{CodeBERT} and \texttt{code-davinci-002}. For \texttt{CodeBERT}, we add a linear classification head to its BERT trunk in order to build a multi-class classification model.  
The inputs to the model are the context and vulnerable code block, separated by a special [SEP] token and bounded by a beginning of sequence [BOS] token and end of sequence [EOS] token
\begin{align*}
    [BOS], c_1, c_2, ...c_n, [SEP], v_1, v_2, ...v_m, [EOS]
\end{align*}
where $c_1, c_2, ...c_n$ denotes a sequence of n tokens of context code surrounding the vulnerability and $v_1, v_2, ...v_m$ denotes a sequence of $m$ tokens of the vulnerable block of code. We distinguish between context and vulnerable block to enable the model to process any given complete or incomplete code snippet.
Because our data is vastly imbalanced with less than 10\% vulnerable examples, we employ an oversampling technique on the vulnerable examples while training: for each epoch, all vulnerable examples are retained while the non-vulnerable examples are rotated through so that each epoch contains 50\% vulnerable and 50\% non vulnerable examples. The model is trained with a standard binary cross entropy (BCE) loss. In the rest of this work, we refer to this model as \textbf{DeepDevVuln}.

Our second fine-tuned model is a fine-tuned version of \texttt{code\-davinci-002} on 30,000 randomly sampled examples from our training set. We down-size our training data due to the cost of fine-tuning the model on the entirety of the training set. In this case, the context and vulnerable block are concatenated together, then a special classification token is appended to the end of the sequence. Because \texttt{code-davinci-002} is a GPT decoder model that does not inherently perform classification, we use next-token prediction as a proxy, i.e. the output is a special token that signifies vulnerable or not. The sequence is as follows:
\begin{align*}
    [BOS], c_1, c_2, ...c_n, v_1, v_2, ..., v_n [CLS] [VULN]
\end{align*}
We refer to this variant as \textbf{CodexVuln}. 

\begin{table}[h]
    \caption{Summary statistics of vulnerable issues gathered from Github PRs}
    \centering
    \begin{tabular}{c|c|c|c}
        Vulnerability & CWE & N \\
        \hline 
        SQL Injection & 89 & 45\\
        Hardcoded Credentials & 798 & 23 \\
        Code Injection& 94 & 13 \\
        Path Injection & 22 & 7 \\
        Clear Text Logging & 312 & 5 \\
        Weak Cryptographic Algorithm & 327 & 5 \\
        Incomplete URL Substring Sanitization & 20 & 2 \\
    \end{tabular}
    \label{table:github_cwes}
\end{table}
\subsection{Model Variant Evaluation}
To better understand the effect of model architecture and training choices, we compare our six model variants on a dataset we collect from GitHub pull requests.

\subsubsection{Metrics}
The majority of existing approaches treat vulnerability detection as a classification problem. Given a block of code, the model makes a binary decision on whether the code is vulnerable or not. Therefore, in our evaluations we also use the most common evaluation metrics for a classification task \cite{powers2020evaluation} which includes: 
\begin{itemize}
    \item \textbf{Precision}: which indicates the correctness of predictions and calculated as true positive / (true positive + false positive)
    \item \textbf{Recall}: which indicates the effectiveness of the prediction and calculated as true positive / (true positive + false negative).
    \item \textbf{F1-score}: which indicates balance between precision and recall and is defined as the geometric mean of the two.
\end{itemize}

\begin{table}[htbp]
    \centering
    \caption{Performance of DeepDevVuln model on Github PR Vulnerabilities dataset.}
    \label{table:github_cwes_results}
    \begin{tabular}{l c c c}
        Model & Precision & Recall & F1-Score \\\hline
        DeepDevVuln & 58.87\% & 63.00\%  & \textbf{60.87}\% \\
        CodexVuln & \textbf{69.56}\% & 48.00\%  & 56.80\% \\
        CodexZero & 11.08\% & \textbf{98.00}\%  & 19.90\% \\
        TextZero & 46.99\% & 78.00\%  & 58.65\% \\
        CodexFew (8 examples) & 23.91\% & 95.00\%  & 37.70\% \\
        TextFew (6 examples) & 49.01\% & 75.00\%  & 59.29\% \\\hline
    \end{tabular}
\end{table}
\subsubsection{Dataset}
To create this dataset, we collect all pull requests that contained \textit{"fix <issue>"} in the pull request title for every combination of issue from Table \ref{table:github_cwes} and language (i.e. JavaScript, Python, Go, Java, C++, C\#, Ruby). To ensure that the retrieved pull requests were relevant to the target vulnerability, we only include the pull requests that contain both removal and addition of code and were less than 100 lines in length. This yields 283 pull requests. We then manually examine each pull request to ensure the vulnerabilities are legitimate.

Using this process, we gather a set of 100 vulnerable examples constituting the CWEs in Table \ref{table:github_cwes}. We added a set of 906 non vulnerable examples to this dataset. To collect the non vulnerable examples, we randomly sample 150 files from repositories that CodeQL has scanned and has not detected an issue. We then randomly select chunks of non-overlapping code between 1 to 10 lines from these files to generate nonvulnerable examples. In total, the GitHub PR dataset contains 1,006 examples.

\begin{figure}[ht]
    \includegraphics[scale=0.5]{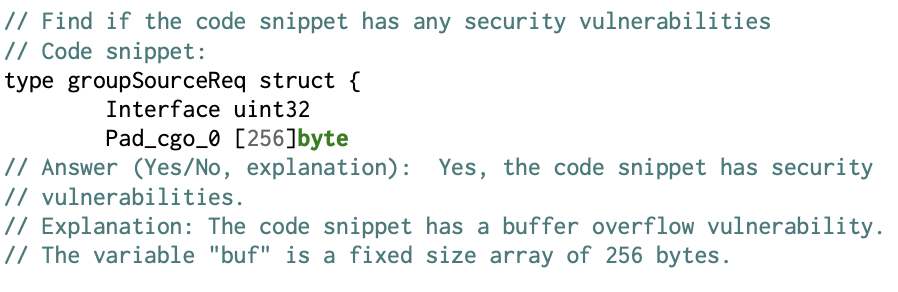}
    \caption{A sample of CodexZero false positive due to overreach}
    \label{fig:codexzero_fp1}
    \end{figure}


\begin{figure}[ht]
    \includegraphics[scale=0.5]{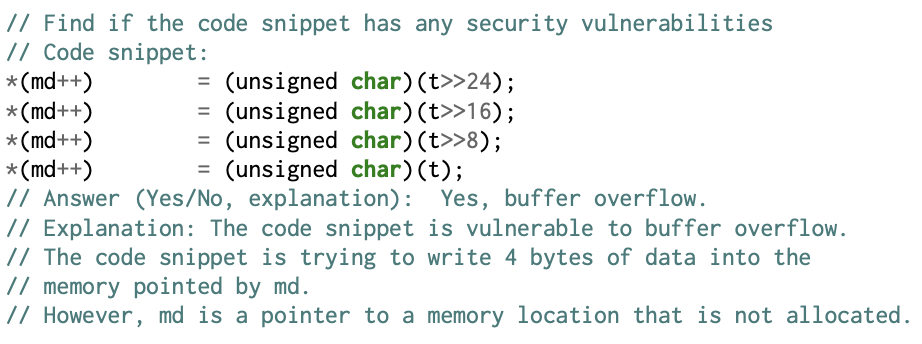}
    \caption{A sample of CodexVuln false positive due to lack of context}
    \label{fig:codexzero_fp2}
    \end{figure}

\subsubsection{Results}
Table \ref{table:github_cwes_results} shows the performance of our model variants on this dataset. For these six model variants, we can see that DeepDevVuln has the best performance with respect to F1 score. Every zero-shot and few-shot variant outperforms DeepDevVuln in terms of recall, however the precision of each variant is significantly lower. One reason might be that the zero-shot and few-shot models have a wider definition of what constitutes vulnerable code. For example, the code snippet presented in Figure \ref{fig:codexzero_fp1} below does not have any explicit vulnerabilities, however, CodexZero considers the worst-case scenario that the code may be vulnerable to a buffer overflow in the future. Due to this tendency, it seems that CodexZero will raise an alert even if there is no explicit vulnerability present, which leads to a high recall score but low precision score.

Another example further explains CodexZero's high false positive rate. In Figure \ref{fig:codexzero_fp2}, we see that the model lacks context around the variable $md$, and explains that because it was never initialized, accessing the memory at that location could lead to a vulnerability.

This explanation extends, in part, to TextZero, which has a more balanced precision and recall. In TextZero's case, the prompt variation used had a significant effect: for example, the first prompt variation \textit{"identify potential security vulnerabilities"} had similar results to CodexZero. However, using the phrase \textit{"detect any security risks"} led to the result in Table \ref{table:github_cwes_results}. This may have encouraged the model to focus on the exact code snippet as-is, rather than to speculate about possible vulnerabilities.

Comparing few-shot and zero-shot results, we see the few-shot evaluations surpassed their zero-shot counterparts. One explanation is that the inclusion of examples gives the model a sense of what to expect in terms of code snippet length and area of focus. For instance, when the non-vulnerable examples include uninitialized variables and incomplete context, the model starts to ignore the worst-case scenarios explained above.

\begin{figure}[ht]
\includegraphics[scale=0.5]{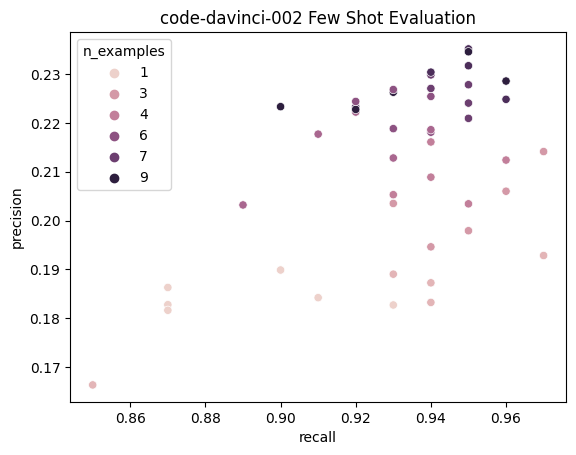}
\caption{Trials varying examples of CodexFew}
\label{fig:codex-few-shot-trials}
\end{figure}

Finding the best number of examples was a matter of gradually increasing the number of examples prepended to the prompt from 1 to 9. We ran several trials and reported the best results. Among these trials, we observed that, especially for CodexFew, there is an apparent correlation between larger number of examples and the precision and recall achieved as seen in Figure \ref{fig:codex-few-shot-trials}. It may be that the exact types of examples selected have a major role in steering the model's output. A similar trend for precision, but not recall, was observed for TextFew, Figure \ref{fig:text-few-shot-trials}.

\section{Experiments}
Below, we explain our experiments designed to answer the following two research questions:
\begin{itemize}
    \item \textbf{RQ1:} How effective is our vulnerability detection model compared to results from state of the art models on established benchmark datasets?
    \item \textbf{RQ2:} To what extent our proposed vulnerability detection model is effective in reducing the vulnerability rate of code LLMs?
\end{itemize}
We perform two experiments to answer these questions. The first experiment focuses on comparing our model against the state of the art vulnerability detection models on common datasets. 
The second experiment gauges the effectiveness of our model in reducing the vulnerability rate of code LLMs.

\begin{table*}[t]
    \caption{Summary of common vulnerability detection datasets.}
    \label{table:datasets}
    \begin{tabular}{l c c c c p{0.4\linewidth}}
        Dataset & \# Programs & \% Vuln &  \# Duplicates & Granularity & Description \\\hline\hline
        VulDeePecker~\cite{li2018vuldeepecker} & 61,638 & 28.76\% & 33,786 & Slice &  {It was obtained by preprocessing examples from the National Vulnerability Database (NVD) and the Software Assurance Reference Dataset (SARD) and consists of CWE-119 (Improper Restriction of Operations within the Bounds of a Memory Buffer) and CWE-399 (Resource Management Errors).} \\\hline
        SeVC~\cite{li2021sysevr} & 420,627 & 13.41\% & 188,030 & Slice &  {An improvement over the VulDeePecker dataset, covering 126 different types of vulnerabilities and divided into four categories: API Function Call, Arithmetic Expression, Array Usage, and Pointer Usage.} \\\hline
        ReVeal~\cite{chakraborty2021deep} & 22,734 & 9.85\% & 351 & Function &  {It tracks past vulnerabilities from the Linux Debian Kernel and Chromium projects. The dataset reflects real bug reports and has a realistic data imbalance, with only 10\% of the examples being vulnerable.} \\\hline
        FFMPeg+Qemu~\cite{zhou2019devign} & 27,318 & 45.61\% & 60 & Function &  {It consists of past vulnerabilities and their fixes from two open-source projects.} \\
        \hline
    \end{tabular}
\end{table*}

\subsection{Experiment 1: Performance against Existing Approaches on Benchmark Datasets}
In this section, we present the experimental evaluation of our vulnerability detection model against existing state of the art approaches on four widely used datasets. Table \ref{table:datasets} summarizes the datasets used in this experiment.

\begin{figure}[ht]
\includegraphics[scale=0.5]{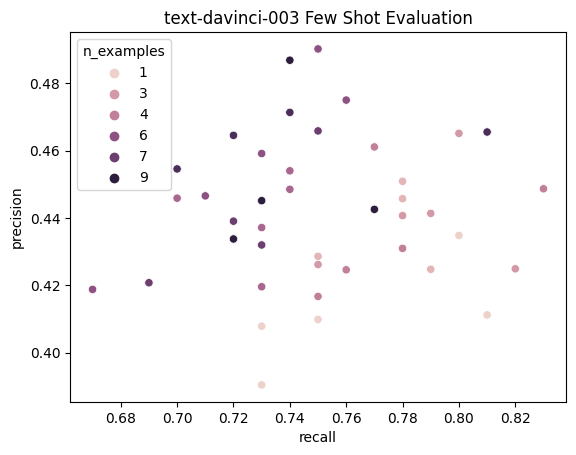}
\caption{Trials varying examples of TextFew}
\label{fig:text-few-shot-trials}
\end{figure}

\subsubsection{Experimental Setup}
The granularity of the data, quality of the data, and types of issues covered in each of the datasets in Table~\ref{table:datasets} is different from our training and test set. Therefore, we followed the approach of Chakraborty et al. \cite{chakraborty2021deep}, where we first remove duplicate examples from each of the datasets. We then fine-tune DeepDevVuln as a binary classification model on each dataset for 10 epochs. For each dataset, we use a standard 80\%/10\%/10\% train/validation/test split, consistent with the baseline models.

\subsubsection{Results}

\begin{table}[htbp]
    \centering
    \scriptsize
    \caption{Performance of DeepDevVuln model on Vuldeepecker, SeVC, Reveal, and FFMPeg+Qemu datasets.}
    \label{table:soa-results}
    \begin{tabular}{l | l | c c c}
        Dataset & Model & Precision & Recall & F1-Score \\\hline\hline
        \multirow{3}{4em}{VulDee-Pecker CWE 119} &  VulDeePecker & 82.00\% &  91.70\% & 86.6\%\\
        & Thapa et al. CodeBERT  & 95.27\% & 95.15\% & 95.21\%\\
        & Thapa et al. GPT-2 Base & 93.35\% & 93.56\% & 93.45\% \\
        & Thapa et al. GPT2-Large & 95.74\% & 95.28\% & 95.51\% \\
        & Codex & \textbf{97.45}\% & 93.31\% & 95.33\%\\
        & DeepDevVuln & 96.74\% & \textbf{95.62}\% & \textbf{96.18}\%\\\hline
        \multirow{3}{4em}{VulDee-Pecker CWE 399} &  VulDeePecker & 95.30\% &  94.60\% & 86.6\%\\
        & Thapa et al. CodeBERT  & 94.25\% & 95.29\% & 94.76\%\\
        & Thapa et al. GPT-2 Base & 92.97\% & 94.99\% & 93.96\% \\
        & Thapa et al. GPT2-Large & \textbf{96.79}\% & 96.90\% & 96.84\% \\
        & Codex & 96.69\% & 97.04\% & \textbf{96.87}\%\\
        & DeepDevVuln & 95.65\% & \textbf{97.41\%} & 96.53\%\\\hline
        \multirow{3}{4em}{SeVC} & Thapa et al. (BERTBase) & 88.73\% & 87.95\% & 88.34\%\\
        & Thapa et al. (GPT-2 Base) & 86.88\% & 87.47\% & 88.34\%\\
        & Codex & 82.26\% & 84.34\% & 83.29\%\\
        & DeepDevVuln & \textbf{95.56\%} & \textbf{97.14\%}  & \textbf{96.35\%} \\\hline
        \multirow{3}{4em}{ReVeal} 
        & Chakraborty et al.  & 30.91\% & 60.91\% & 41.25\%\\
        & Codex & 45.04\% & 29.80\% & 35.87\%\\
        & CodeBERT & \textbf{48.95\%} & 35.35\% & 41.06\% \\
        & DeepDevVuln & 41.00\% & \textbf{61.00\%} & \textbf{49.29\%} \\\hline
        \multirow{3}{4em}{FFmpeg + Qemu} 
        & Chakraborty et al. & 56.85\% & 74.61\% & 64.42\%\\
        & Codex & \textbf{63.22\%} & 55.64\% & 59.19\%\\
        & CodeBERT & 62.94\% & 58.70\% & 60.74\% \\
        & DeepDevVuln & 57.34\% & \textbf{78.06\%} & \textbf{66.11\%} \\\hline
    \end{tabular}
\end{table}

As shown in Table \ref{table:soa-results}, our DeepDevVuln model has the best overall F1-Score for the majority of datasets. This means that our model demonstrates a good balance between precision and recall which is important for vulnerability detection. Additionally, our vulnerability detection model has 10 times fewer parameters than GPT2-Large or Codex, yet still achieves comparable precision and higher recall. Overall this results suggests that our approach allowed our model to adapt to the specific types of issues present in these datasets and leverage its knowledge gained through pretraining on our vulnerability dataset to improve upon the state of the art results.

\subsection{Experiment 2: Model's effectiveness in reducing the vulnerability rate of code LLMs}
In our second experiment, we evaluated the effectiveness of our vulnerability detection model in detecting the vulnerable code completions of four different code LLMs:
\begin{itemize}
    \item \texttt{CodeGen-2B}: a transformer decoder model trained on natural language and code (C, C++, Go, Java, JavaScript, Python)
    \item \texttt{code-cushman-001}: smaller-size Codex model, trained on source code from GitHub
    \item \texttt{code-davinci-002}: full-size Codex model, trained on source code from GitHub
    \item \texttt{text-davinci-003}: Codex model based on InstructGPT \cite{instructgpt}, using reinforcement learning from human feedback (RLHF)
\end{itemize}

We evaluated the extent to which our model detects the vulnerable code patterns produced by each LLM utilizing the benchmark created by Pearce et al\cite{nyu_benchmark}. This benchmark consists of scenarios to evaluate the likelihood of a code LLM generating vulnerabilities. These scenarios are constructed to specifically elicit a completion containing a particular vulnerability. Each scenario is associated with a particular CWE and includes a prompt code snippet and a corresponding CodeQL query. The prompt is used as input to a code LLM. The model-generated completion is appended to the prompt and this completed snippet is then analyzed using the provided CodeQL query. Completions that cannot be analyzed by CodeQL (due to syntactical errors) are considered invalid and excluded from the analysis. CodeQL marks the remaining valid completions as either vulnerable or clean.

We took the 29 Python scenarios developed by \cite{nyu_benchmark} and, following the same process, we added 11 new JavaScript scenarios covering 10 CWEs to the benchmark. Table \ref{tab:benchmark} describes the scenarios we added using the same format as in \cite{nyu_benchmark}. ``Rank" reflects the CWE ranking in the MITRE list if applicable. CWE-Scn refers to the scenario's identifier and associated CWE. All of these scenarios are written in JavaScript, originate from the public GitHub CodeQL documentation, and were evaluated with CodeQL.

\begin{table}[h]
    \centering
    \caption{Javascript Scenarios covering 10 CWEs in Javascript}
    \begin{tabular}{l|c|l}
        Rank & CWE-Scn. & Description\\
        \hline \\
        3 & CWE-89 & SQL Injection \\
        4 & CWE-20 & Incomplete Url Substring Sanitization \\
        8 & CWE-22 & Tainted Path \\
        15 & CWE-798 & Hardcoded Credentials\\
        25 & CWE-94 & Code Injection\\
        35 & CWE-601 & Client Side Url Redirection\\
        35 & CWE-601 & Server Side Url Redirection\\
        40 & CWE-312 & Clear Text Storage of Sensitive Data\\
        - & CWE-209 & Stack Trace Exposure\\
        - & CWE-327 & Broken Cryptographic Algorithm\\
        - & CWE-916 & Insufficient Password Hash\\
    \end{tabular}
    \label{tab:benchmark}
\end{table}

For each scenario, a model can generate a varying number of valid or vulnerable completions. In the context of code LLMs, a developer may often see only a single completion for a given prompt. Therefore, we evaluate vulnerability rates at the level of scenarios (prompts): we count the number of scenarios that yielded at least one vulnerable completion. For example, suppose there are 10 scenarios and each model generates 5 completions per scenario. For each of the 10 scenarios, we run the corresponding CodeQL query on its 5 completions. Suppose that 9 scenarios have at least one syntactically valid completion. We consider the 9 scenarios with valid completions and examine how many of the 9 have at least one vulnerable completion.

\newcommand{\specialcell}[2][c]{%
  \begin{tabular}[#1]{@{}c@{}}#2\end{tabular}}
\begin{table*}[ht]
    \centering
     \caption{Vulnerability rate on scenario level for different baselines.}
    \begin{tabular}{l|c|c|c|c|c}
        & \multicolumn{2}{c|}{Without DeepDevVuln} & \multicolumn{2}{c|}{With DeepDevVuln} & \\
        Approach & \specialcell{Valid \\Scenarios} & \specialcell{Vulnerable \\Scenarios} & \specialcell{Valid \\Scenarios} & \specialcell{Vulnerable \\Scenarios} & \specialcell{Vulnerability \\Reduction Rate}\\
        \hline 
        CodeGen (6B) & 7 & 7 (100.00\%) & 7 & 2 (29.00\%) & 71.00\% \\
        code-cushman-001 & 25 & 25 (100.00\%)& 19 & 5 (26.0\%) & 74.00\% \\
        code-davinci-002 & 26 & 24 (92.00\%)  & 20 & 7 (35.0\% & 61.96\% \\
        text-davinci-003 & 27 & 21 (78.00\%) & 24 & 2 (8.0\%) & 89.74\% \\
        
    \end{tabular}
    \label{tab:overall-results}
\end{table*}

For each model, we generate 25 completions per scenario. We then run our vulnerability detection model on each completion and filter out the completions that our model detects as vulnerable. We then rerun the CodeQL queries from the benchmark on the remaining completions.

\subsubsection{Results}

The results of this vulnerability experiment are shown in Table \ref{tab:overall-results}. The first two columns corresponds to each code LLM acting alone, while the second two columns includes filtration from our vulnerability detection model. The vulnerability reduction rate is the percentage reduction in the vulnerability rate as a result of filtration.

As shown in the table, filtering vulnerable outputs results in a significant decrease in vulnerability rate. In particular, the vulnerability reduction rate is highest for \texttt{text-davinci-003}, which follows InstructGPT's method of reinforcement learning from human feedback (RLHF). RLHF is known to substantially improve the quality and naturalness of generated text. Therefore, \texttt{text-davinci-003} likely generates code that more closely resembles code written by real developers. Since DeepDevVuln is trained on developer-written code, it may be better able to detect vulnerabilities in outputs from \texttt{text-davinci-003} than other code LLMs.

\section{Deploying in Production}
We deployed our model for detecting vulnerable code patterns in a VSCode extension with $\sim$100K daily users. After each key stroke that a user writes, their incomplete code snippets are sent to our model for verification. To evaluate the effectiveness of our model, we collected and examined the JavaScript code snippets that were sent to our model in a three month period, from November 2022 to January 2023, for a total of $\sim$6.7M code snippets. We chose to focus on JavaScript because it is the most popular language in VSCode. 

In order to have a baseline for comparison, we ran all the CodeQL security queries for JavaScript on the collected code snippets. Overall, CodeQL detected $\sim$1,284 vulnerable code snippets. However, this number is a lower-bound for the amount of actual vulnerable code snippets. CodeQL queries do not successfully run and detect issues in all of the code snippets, because CodeQL is designed to run on a set of complete files in a repository. Therefore, CodeQL's vulnerability detection rate drops significantly when executed on syntactically-incorrect code or incomplete code that is presented in a single file as opposed to the repository context. This drop impacts some scenarios more than others, depending on sensitivity of the query to syntax issues and the amount of context required by the query to detect a vulnerability. Of CodeQL's vulnerability detections, over 58\% were from two scenarios which have simple CodeQL queries requiring less context to run successfully.  In comparison, DeepDevVuln detections were more uniform. In fact, over 57\% of DeepDevVuln detections were from SQL Injection, Code Injection, Client Side URL Redirect, Server Side URL Redirect, and Insufficient Password Hash scenarios. This is significant because JavaScript is a dominant language in both client- and server-side web development, and these classes should be more prominent in this domain. Yet, CodeQL detects these scenarios at a rate of less than 1 in 1,000,000.  CodeQL's poor coverage and inability to detect vulnerabilities on this production data highlights the need for deep learning based detection systems in live services.

\begin{figure}[h]
    \centering
    \includegraphics[width=0.4\textwidth]{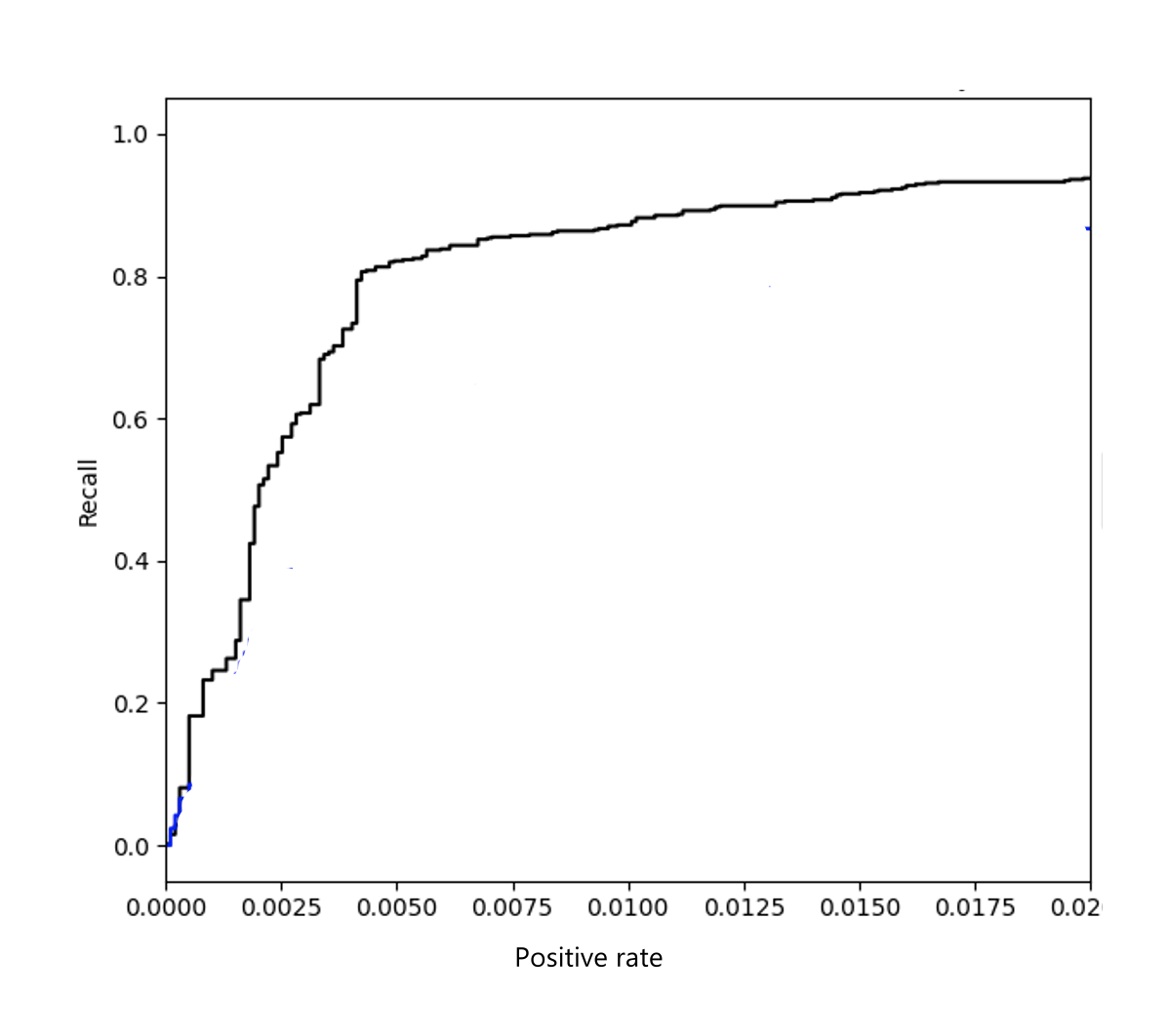}
    \caption{DeepDevVuln model's performance on recall vs. positive rate}
    \label{fig:recall-filteration}
\end{figure}

For our evaluation, because CodeQL-detected issues are a lower-bound on the number of issues, we use them to measure recall. Instead of precision, we measure the positive rate (number of detected issues over number of all scanned issues). Figure \ref{fig:recall-filteration} shows how our DeepDevVuln model performs on recall vs positive rate: it can achieve up to 90\% recall with around 1\% positive rate. While we optimized for recall for our extension, other applications can find the right balance between recall and positive rate based on their user scenario and feedback.  

Overall, we observed that DeepDevVuln's vulnerability reduction rate (i.e. the reduction in the rate of vulnerabilities present in developer's code) for JavaScript was 89.64\%.

\section{Lessons Learned and Ongoing Work}
\label{discussion}
In the process of building and deploying our model, we learned a few lessons that have guided our ongoing and future work on vulnerability detection.

First, the different learning methods we explored in this work come with trade-offs in model size, inference cost, predictive performance, and maintenance cost. Zero-shot and few-shot learning require sufficiently large models in order to effectively make predictions. Furthermore, predictive performance tends to improve with model size. For a constant model size, the inference cost of zero-shot  learning is slightly higher than for fine-tuning, since a prompt must be constructed for each example; the cost for few-shot learning is even higher, since the system must fetch examples for each example. Our results show that a fine-tuning approach yielded the best prediction performance, allowing us to make accurate predictions without incurring high inference cost. However, in order to maintain a fine-tuned model, we must continually monitor and retrain the model on additional vulnerability types. Zero-shot and few-shot learning, on the other hand, would only require maintenance with regard to prompts and examples, rather than any further training.

Second, there is a trade-off between the size of a model and its response time. This work focuses on detecting vulnerabilities at \textit{EditTime}, while a developer is writing code in an IDE. A large vulnerability detection model requires more time to make predictions, which can result in delayed response and the developer missing the vulnerability. In order to maintain low prediction latency, our vulnerability detection is based on the relatively small CodeBERT-base model and has under 100M parameters. As more powerful hardware is built to run large models and improve the inference time, we expect to be able to run large models in production settings in our future iterations.

Third, in many classification problems, a model's prediction threshold is used to create an appropriate balance between precision and recall. The balance is important because an effective production-ready fault detector must minimize churn and false positives \cite{bessey2010few}. High churn, where the issues raised vary from one version of the system to another, can cause significant friction with users due to a lack of consistency. False positives can similarly erode developer trust in the usefulness of a system, resulting in developers ignoring the tool. In our case, we do not have the ground truth of all vulnerable code snippets for our live evaluations, and therefore we cannot measure precision In this case, the analogous metrics are positive rate (the fraction of examples that the model predicts as positive) and recall. Our second lesson was in balancing these metrics for a production-scale vulnerability detection system. 
We tuned our threshold to maintain a 1\% positive rate based on initial user's interactions and feedback. However more long-term studies and monitoring of these metrics are needed to better adjust the balance.

Finally we learned that we should periodically retrain our model to expand coverage as new vulnerability types are caught. When the common vulnerabilities are caught early on in the development process, users may start to notice the uncommon vulnerabilities and this may hurt their trust on detection tools overtime. Therefore, to address this challenge we have implemented a retraining pipeline where we constantly find datasets with new vulnerabilities to feed the pipeline and expand the coverage. 

\section{Conclusion}
 Code vulnerabilities continue to cost software companies and users. Detecting vulnerabilities in code at \textit{EditTime} when the code is written by a developer or generated by a code LLM is essential to ensure the vulnerabilities are fixed at lower cost. Yet, the majority of current vulnerability detection tools do not detect vulnerabilities at \textit{EditTime}. Our work closes this gap by presenting a vulnerability detection model that detects vulnerabilities on incomplete code snippets and therefore can be used to detect vulnerable code patterns at \textit{EditTime} when code LLMs or developers write them. Our evaluation results showed that our model improves the state of the art detection approaches by 10\% in recall and 8\% in precision. In addition, our model reduces the vulnerability rate of code LLMs by $>$89\%.  

An immediate direction for future work is to expand our vulnerability detection model's coverage with adding new types of vulnerabilities to our training set. Another direction is to measure the long-term effect of our vulnerability detection model on the overall experience of developers who are using our VSCode extension. For example, measures can include the vulnerability reduction rate, whether the file under development resulted in a unit test failure, or whether a vulnerability was caught in the file after the \textit{EditTime} (e.g. build time). 



\bibliographystyle{ACM-Reference-Format}
\bibliography{references}


\end{document}